\RequirePackage[hyphens]{url} 
\documentclass[10pt, a4paper]{ieeeconf}      
\IEEEoverridecommandlockouts                              
\usepackage[flushleft]{threeparttable}
\usepackage{url}
\usepackage{times}
\usepackage{url}
\usepackage{color}
\usepackage{gensymb}

\PassOptionsToPackage{hyphens}{url}\usepackage{hyperref}

\newcounter{comment}

\usepackage[noadjust]{cite}
\usepackage[listings,skins]{tcolorbox}
\usepackage{listings}
\usepackage{pxfonts}
\usepackage{courier}
\usepackage[colorinlistoftodos]{todonotes}
\usepackage{apacite}

\newcounter{mycomment}

\usepackage{fancyhdr}
 
\title{Ambitious Data Science Can Be Painless$^{*}$}

\author{Hatef Monajemi$^{1,2}$, Riccardo Murri$^{3}$, Eric Jonas$^{4}$, Percy Liang$^{5}$, Victoria Stodden$^{6}$ and David Donoho$^{\dagger,1}$\\
\vspace{2em}
\today
\thanks{* This article is based on a series of lectures given in a Stanford course Stats285 in the Fall of 2017.}
\thanks{$^{\dagger}$ Corresponding Author,
        {\tt\small donoho@stanford.edu}}%
\thanks{$^{1}$ Dept. of Statistics, Stanford University}%
\thanks{$^{2}$ Data Science Initiative, Stanford University}%
\thanks{$^{3}$ S3IT, University of Zurich}%
\thanks{$^{4}$ Dept. of Computer Science, University of California Berkeley }%
\thanks{$^{5}$ Dept. of Computer Science, Stanford University }%
\thanks{$^{6}$ School of Information, University of Illinois Urbana-Champaign }%
}

\begin{document}

\maketitle
\thispagestyle{fancy}
\pagestyle{plain}



\begin{abstract}
Modern data science research, at the cutting edge, can involve massive computational experimentation; an ambitious PhD in computational fields may conduct experiments consuming several million CPU hours. Traditional computing practices, in which researchers use laptops, PCs, or  campus-resident resources with shared policies, are awkward or inadequate for experiments at the massive scale and varied scope that we now see in the most ambitious data science.
On the other hand, modern cloud computing promises seemingly unlimited computational resources that can be custom configured, and seems to offer a powerful new venue for ambitious data-driven science. {Exploiting the cloud fully, it seems the amount of raw experimental work that could be completed in a fixed amount of calendar time ought to expand by several orders of magnitude.}

As potentially powerful as cloud-based experimentation may be in the abstract, it has not yet become a standard option for researchers in many academic disciplines. The prospect of actually conducting massive computational experiments in today's cloud systems with today's standard approaches forcefully confronts the potential user with daunting challenges.  The user schooled in traditional interactive personal computing likely expects that a cloud experiment will involve an intricate collection of moving parts seemingly requiring extensive monitoring and involvement.  Leading considerations include: \emph{(i)} the seeming complexity of today's cloud computing interface, \emph{(ii)} the difficulty of executing and managing an overwhelmingly large number of computational jobs, and \emph{(iii)} the difficulty of keeping track of, collating, and combining a massive collection of separate results. Starting a massive experiment `bare-handed' seems therefore highly problematic and prone to rapid `researcher burn out'.

New software stacks are emerging that render massive cloud-based experiments relatively painless. Such stacks simplify experimentation by systematizing experiment definition, automating distribution and management of all tasks, and allowing easy harvesting of results and documentation.
{In this article, we discuss several \emph{painless computing stacks} that abstract away the difficulties of massive experimentation, thereby allowing a proliferation of ambitious experiments for scientific discovery.
}
\end{abstract}



\begin{table*}
\caption{Features of the Painless Computing Stacks Presented in This Paper$^2$}
\centering

\begin{threeparttable}
\begin{tabular}{l|ccc}
\hline
 & ElastiCluster/ClusterJob & CodaLab & PyWren \\
\hline
Scientific Service Layer & Interface & Interface & Framework \\
Service Type & EMS & EMS & Serverless Execution\\
Input & Script & Script & Function/Values\\
Language & Python/R/Matlab & Agnostic & Python \\
Server Provisioning & Yes & Yes & Automatic\\
Resource Manager & SLURM/SGE & Agnostic & Automatic \\
Job Submission & Yes &  Yes & Yes \\
Job Management & Yes & Yes & Automatic \\
Auto Script Parallelization\tnote{1}  & Yes & No & No \\
Auto Mimic & No & Yes & No \\
Auto Storage & No & Yes & No \\
Experiment Documentation & Yes & Yes & No\\
Reproducible & Yes &  Yes & Yes \\
\hline
\end{tabular}
\begin{tablenotes}
\item[1]{For embarrassingly parallel scripts only.}
\item[2]{This table includes available features as of December 2018.}
\end{tablenotes}
\end{threeparttable}
\label{tbl_feature_model}
\end{table*}

\section{Introduction}

Tremendous increases in computing power in  recent years are opening fundamentally new opportunities in science and engineering. Amazon, IBM, Microsoft and Google now make massive and versatile compute resources available on demand via their \emph{cloud} infrastructure, making it in principle possible for a broad audience of researchers to individually conduct ambitious computational experiments consuming millions of CPU hours within calendar time scales of days or weeks. We anticipate the emergence of widespread massive computational experimentation as a fundamental avenue towards scientific progress, complementing traditional avenues of induction (in observational sciences) and deduction (in mathematical sciences) \cite{MMCEP17,fourth-paradigm}. Indeed, there have been recent calls by funding agencies (e.g.,  NSF \cite{NSFCloudCall},  NIH \cite{NIHStrategicPlan,Strides18}) for the greater adoption of cloud computing and Massive Computational Experiments (MCEs) in scientific research.

In some fields, this emergence is already quite pronounced.
The current remarkable wave of enthusiasm for machine learning (and its deep learning variety)
seems, to us, evidence that massive computational experimentation has
begun to pay off, big time.  Deep neural networks have
been around for the better part of 30 years;  
but only in recent years
have they been able to successfully penetrate in certain applications.
What changed recently
is that researchers at the cutting edge can experiment
extensively with tuning such nets and refactoring them;
with enough experimentation,
dramatic improvements over older methods have been found, thereby changing the game.

Indeed, experimental success has disrupted field after field.
In machine translation, many major players,
including Google and Microsoft, moved away recently from Statistical Machine Translation (SMT) to Neural Machine Translation (NMT) \cite{Microsoft-MT, Google-MT}.

Similar trends can be found in 
computer vision \cite{AlexNet12,vgg14}, and many other areas of artificial intelligence. 
Tesla Motors is now using predominantly deep neural networks in their decision-making systems, according to Andrej Karpathy, the head of their AI department\footnote{Source: a public lecture titled ``Software 2.0'' by Karpathy, at Stanford's Computer Science depertment on January 17, 2018.}. 

In a nutshell, if in the past the answer to ``how to improve task accuracy'' was to ``use a better mathematical model for the situation'', today's answer seems to be ``exploit a bigger database'' and ``experiment with different approaches until you find a better way.''\footnote{Even in Academic Psychology!  \cite{Yarkoni17} }  Evidently, this shift towards adopting computational experiments for problem solving is tightly linked to the explosion of computational resources in  recent years.\footnote{A recent analysis by OpenAI shows that the amount of compute used in modern AI systems grew exponentially since 2012, doubling each 3.5 months \cite{OpenAI}.}    



More generally, massive experimentation can solve complex problems lying beyond the reach of any theory.
Successful examples of ambitious computational experimentation as a fundamental method of scientific discovery abound:
in \cite{Brunton16}, the authors take a data science approach to discover governing equations of various dynamical systems including the strongly nonlinear Lorenz-63 model; in \cite{Monajemi13} and \cite{MoDo18}, the authors conducted data science studies involving several million CPU hours to discover fundamentally more practical sensing methods in the area of Compressed Sensing; in \cite{huang15}, MCEs solved a 30-year-old puzzle in the design of a particular protein.


In the emerging paradigm for ambitious data science, scientists pose {bold} research questions to settle via MCEs, followed by careful statistical analysis of data and inductive reasoning \cite{donoho-50-years,tukey-foda,Berman2018}. Under this paradigm researchers may launch and manage historically unprecedented numbers of computational jobs (e.g. possibly even millions of jobs). Users trained in an older paradigm of interactive `personal' computing may expect operations
at such a massive scale to be infeasible, anticipating a lengthy and painful process involving many moving parts and manual interactions with complex cloud services. Conducting computations at such massive scale can thus be perceived as an insurmountable obstacle by many experienced researchers, who otherwise might naturally design and conduct MCEs for scientific discovery. This paper presents several emerging stacks that minimize the difficulties of conducting MCEs using the cloud. These stacks offer high-level support for MCEs, masking almost all of the low-level computational and storage details.

\section{Science In The Cloud}
We have argued that today's most ambitious data science 
studies have the potential to be computationally very demanding, often involving millions of CPU hours.
Traditional computing approaches, in which researchers use their personal computers or campus-wide \emph{shared} HPC clusters can be inadequate for such ambitious studies: laptop and desktop computers simply cannot provide the necessary computing power; shared HPC clusters, which are still the dominant paradigm for computational resources in academia today, are becoming more and more limiting because of the mismatch between the variety and volume of computational demands, and the inherent inflexibility of provisioning compute resources governed by capital expenditures. For example, Deep Learning (DL) researchers who depend on fixed departmental or campus-level clusters face a serious obstacle: today's DL demands heavy use of GPU accelerators; researchers can find themselves waiting for days to get access to such resources, as GPUs are rare commodities on many general-purpose clusters at present. 

In addition, shared HPC clusters are subject to \emph{fixed policies} while different projects may have completely different (and even conflicting) requirements. As an example, consider two different types of experiments: \emph{(i)} an embarrassingly-parallel\footnote{See definition in Section~\ref{sec:transparent-invocation-model}.} experiment characterized by many short-lived jobs that produce a large number of small files, and \emph{(ii)} a large MPI job that runs for a week to produce a small number of very large files. These two experiments have different technical requirements in terms of network latency, memory, storage, software and even scheduler configuration; however, when building and configuring an HPC cluster, a decision must be taken over a class of \emph{target compute tasks}: subsequently, cluster hardware is bought, and scheduler policies are set and enforced, according to this decision.  Batch jobs whose characteristics are close to these target conditions will find an optimal environment, whereas other types of computational tasks will be penalized in terms of turnaround time or other policy-determined limitations. As a specific example, consider low-latency high-speed networking, which is still expensive: organizations providing clusters with such a network would naturally want to maximize the return of their investment; as a result, they set a policy that incentivizes tightly-coupled parallel jobs, which can make heavy use of the low-latency network. This policy then penalizes ``embarassingly parallel'' workloads and disappoints the users who run this kind of experiments.

On the other hand, the advent of cloud computing offers instant on-demand access to \emph{virtually infinite computational resources} that can be custom configured to satisfy the needs of individual research projects. Google Cloud Platform (GCP), Amazon Web Services (AWS), Microsoft Azure and other cloud provides now offer easy access to a large array of virtual machines that cost pennies per CPU hour, making it possible for individual research groups to perform 1 Million CPU hours of computation over a few calendar days at a retail cost of perhaps ten thousand dollars. The cloud providers also offer access to many GPUs for as low as 45 cents/hour,  making them an affordable medium for Deep Learning research.

The cloud thus offers several advantages over traditional HPC clusters:
\begin{itemize}
\item \emph{Scalability and Speed}. With millions of servers spread across the globe, cloud providers today own the biggest computing infrastructures in the world. Therefore, any research group with sufficient research funding can \emph{almost instantly} scale out its computational tasks to thousands of cores without having to wait in a long queue on a shared HPC cluster.
\item \emph{Flexibility}. Researchers can adjust the number and configuration of compute nodes depending on their individual computational needs.
\item \emph{Reliability}. Public cloud infrastructures have been
  initially built by large IT companies for their own needs and are
  now relied upon by millions of businesses all over the world for
  their daily computing needs --- they are thus monitored
  $24 \times 7$ and offer excellent uptime and reliability. A good
  example is Netflix that now operates fully on AWS. 
  In fact, Netflix originally decided to migrate entirely to AWS   because the cloud offered
  a more reliable infrastructure \cite{netflix-migration}.
\end{itemize}

%
%
%
%
%
%

Despite massive use of the cloud by business, and the cloud's great potential for hosting ambitious computational studies, many academic institutions and research groups have not yet widely adopted the cloud as a computational resource and continue to use personal computers or in-house shared HPC clusters. We believe that much of the in-house computing inertia is due to the perceived complexity of doing massive computational experiments in the cloud. Users schooled in the interactive personal computing model that dominated academic computing in the 1990-2010 period are psychologically
prepared to see computing as a very hands-on process.
This hands-on viewpoint is likely to perceive a large
computing experiment in terms of the many
underlying individual computers, file systems, 
management layers, files, and processes.  
Users coming from that background may expect 
MCEs to require raw unassisted manual interaction with these moving parts, and would probably anticipate that such manual interaction would be very problematic to complete, as there could be many missteps and misconfigurations in carrying out such a complex procedure. 

If, truly, the cloud-based experiments involved such manual interaction, the process would at best be exhausting and at worst painful.
The many possible problems that could crop up in managing processes manually at the indicated scale would likely be experienced as an overwhelming drag, sapping the experimenter's desire to persevere. Even once the experiment was completed, the burden of having conducted it would likely cast a longer shadow, having drained the analyst's energy and clarity of purpose, thereby reducing the analyst's ability to think clearly about the results.

Summing up, such negative perspectives on cloud-based experiments stem from: \emph{(i)} the perceived complexity of today's cloud computing interfaces, \emph{(ii)} the perceived difficulty of managing an unprecedentedly large number of computational jobs, and \emph{(iii)} the unmet challenge of ensuring that the results of the experiments can be understood and/or reproduced by other independent scientists.

\begin{figure*}
\centering
  \label{fig:layers}
  \includegraphics[width=.8\linewidth]{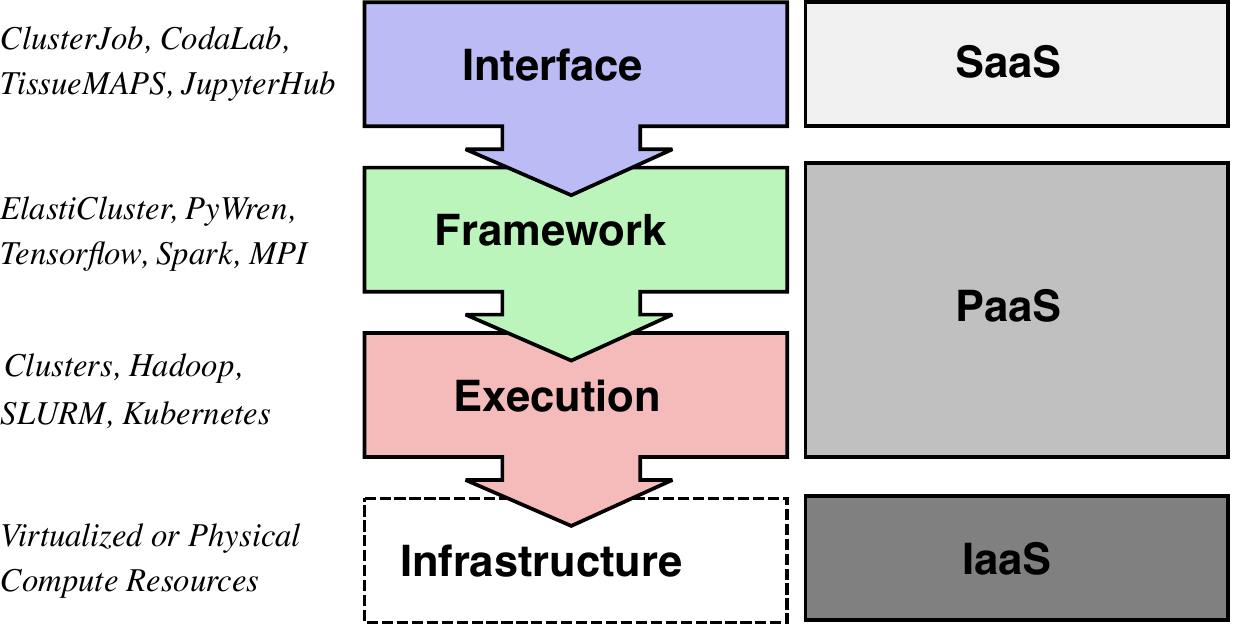}
  \caption{%
    The layering of services for scientific computing in the cloud
    (middle) with some examples (left), compared to the NIST classification of
    cloud-based IT services (right).
  }%
\end{figure*}

\section{The Need for Automation and Painless Computing}

Proper automation of computational research activities \cite{Waltz43} seems 
a compelling way to make massive computational experiments painless and reproducible.
In fact
the vision dates back more than 50 years.

In particular, in his seminal paper ``the future of data analysis''  \cite{tukey-foda}, John Tukey called for the use of automation in data analysis,
arguing, against the critics of automation (see his Section 17) that
\begin{itemize}
\item properly automated tools encourage busy data analysts to study the data more,
\item automation of known procedures provide data analysts with the time and the stimulation to try out new procedures, and
\item it is much easier to intercompare automated procedures.
\end{itemize}

Tukey could not have foreseen the modern context for such automation,
which we now formalize.
As we see it, an ambitious data science study involves:
\begin{enumerate}
\item {\it Precise specification} of an experiment, which includes defining performance metrics and a range of systems to be studied.
\item {\it Distribution, execution, and monitoring} of all the jobs implicitly required in 1).
\item {\it Harvesting} of all the data produced in 2).
\item {\it Analysis} of the data collected in 3).
\item {\it Iterations} of steps (1-4) to run new jobs that may be suggested or required by the results obtained in 4).
\item {\it Reporting} and dissemination of acquired knowledge.
\end{enumerate}

Additionally, the underlying experiment, to be considered ambitious,
may involve either ambitious scale in the data, the computations,
or both.

As must now be apparent, to operate at an ambitious level, it is crucial to automate all these steps and integrate them seamlessly. 
Unlike 50 years ago, automation of data science activities is no longer a \emph{choice} but instead a \emph{necessity}.

In this article, we describe a few examples of software stacks that facilitate such automation; we call them  \emph{Experiment Management Systems} (EMSs). Examples we discuss include
CodaLab Worksheets \cite{CodaLab} and ClusterJob \cite{clusterjob}, which are discussed in detail later in the paper. More generally, we use \emph{Painless Computing Stack} (PCS) to refer to a software stack that abstracts away the difficulties of doing large-scale computation on remote computing infrastructures\footnote{See section \ref{sec:taxonomy} for a finer-grained classification of these systems.}.

As we have already argued in the previous section, unassisted cluster computing (i.e. without EMS assistance) would indeed be painful and draining.
Consider the scientist wanting to spread an ambitious workload
across multiple shared clusters available via XSEDE\footnote{https://www.xsede.org/ecosystem/resources} ecosystem.
We can envision the scientist using traditional practices quickly becoming frustrated with differences in policies, software environment, choice of scheduler, submission rules, licensing differences \cite{RRS}, and other requirements for different clusters.
Refactoring existing properly working single-processor `laptop-scale' scripts
might also be required, imposing an extra unnecessary development
and source code management burden for the scientist.
Finally merely keeping track of progress on each of several different clusters
could be distracting and confusing. Crucially, computational reproducibility is a core requirement of {\it scientific} data science, because the scientific context requires trust in computational findings and safeguards against possible errors. Ensuring that computations done on a cluster can be reproduced at a later time requires additional important considerations often neglected when manual intervention is involved \cite{Donoho2009,Stodden2018,Stodden2016,Berman2018}. Fortunately, the advent of open-source \emph{container} technologies such as Docker \cite{docker14} and Singularity \cite{kurtzer17}, language-agnostic package managers such as Conda \cite{conda}, and common data platforms such as Google's dataCommons (\url{https://datacommons.org}) provides a path for facilitating reproducibility in ambitious cloud experiments.


The common vision motivating the development of the
several PCS's we describe below is
that such draining and ultimately confusing demands be abstracted away. This increase in abstraction 
ought to encourage a proliferation
of ever more ambitious experiments, while
enabling better clarity of interpretation,
better reproducibility, and
ultimately better science.

\section{A Taxonomy of Services for Scientific Computing in the Cloud}
\label{sec:taxonomy}
To better describe how computing stacks presented in the next section interact with cloud infrastructures, we propose a taxonomy of currently available services for
doing scientific computing in the cloud.
The reader should keep in mind that the PCS's presented in this article are not necessarily cloud-based, and can well be used in traditional
on-premises HPC clusters.
We however believe that the coupling
of these systems with cloud infrastructures results in greater advantages for
scientific research by enabling \emph{very} large computational experiments.

In 2011, NIST introduced \cite{nist-definition-of-cloud-computing} a widely-accepted classification of services offered by cloud providers into three layers:
\begin{itemize}
\item \emph{Infrastructure-as-a-Service (IaaS)}: provisioning of compute, storage, networking or other fundamental computing resources.
\item \emph{Platform-as-a-Service (PaaS)}: high-level frameworks and tools to create and run applications on the cloud infrastructure;
\item \emph{Software-as-a-Service (SaaS)}:  end-user applications, whose interface (accessed programmatically or through a web client) is tailored to specific tasks.
\end{itemize}

Scientific computing services typically fall under \emph{PaaS} or \emph{SaaS} in the NIST definition; we will introduce a finer-grained classification applicable to scientific computing applications (see Figure~\ref{fig:layers}):

\begin{enumerate}
\item \emph{Execution layer:}  in our definition, this is the bottom layer and includes services
that can take and run a user-provided program, possibly together with some
specification of the raw computing resources needed at runtime (e.g.,
number of CPU cores, amount of RAM).  Examples are batch-computing clusters,
Hadoop/YARN clusters, container orchestration systems such as Mesos or Kubernetes, and serverless computing services such as AWS Lambda and AWS Batch.
\item \emph{Framework layer:} This layer sits on top of the execution layer and provides users with a way to describe computation in a way that is dictated by an abstract computation model \-- independent of the raw computing resources actually used.  The purpose of the framework layer is to map the abstract computation graph onto a format that can be understood by the execution layer.  Examples
the framework layer are PyWren (see \cite{Jonas17} and Section~\ref{sec:transparent-invocation-model}
later in this paper), Apache Spark
\cite{zaharia2010spark}, TensorFlow
\cite{abadi2016tensorflow},
MPI
\cite{walker1996mpi, mpi-std-3.1}, and
GC3Pie
\cite{maffioletti2012gc3pie,maffioletti12}.

\item \emph{Interface layer:} This layer is the topmost layer in our taxonomy and includes services tailored to a specific set of tasks, masking almost all the details of actual computation and storage management. Examples are ClusterJob \cite{clusterjob} (see Section~\ref{sec:cj}), CodaLab \cite{CodaLab} (see Section~\ref{sec:codalab}), and TissueMAPS (an integrated platform for large-scale microscope image analysis) \cite{tissuemaps-thesis}.

\end{enumerate}

\section{Painless Computing Stacks}
\label{sec:painless-models}

In this section, we present several examples of computing stacks that we consider
relatively pain-free for doing large-scale data science studies in the cloud.

In some cases, these systems permit ambitious experiments that otherwise would be inconceivable to conduct. In some other cases, they \-- without a doubt \-- render experimentation painless, thereby allowing scientists to experiment more. An exact assessment of the extent to which experimentation pain is removed when these stacks are used is beyond the scope of the current article and requires further investigation. There however is ample anecdotal evidence from scientists in different disciplines, which shows a substantial degree of ease and efficiency in experimental research where these tools are exploited.


\subsection{ElastiCluster-ClusterJob Stack}
\label{sec:cj}

\begin{figure}\label{fig:ec-cj}
\centering
\includegraphics[scale=.3]{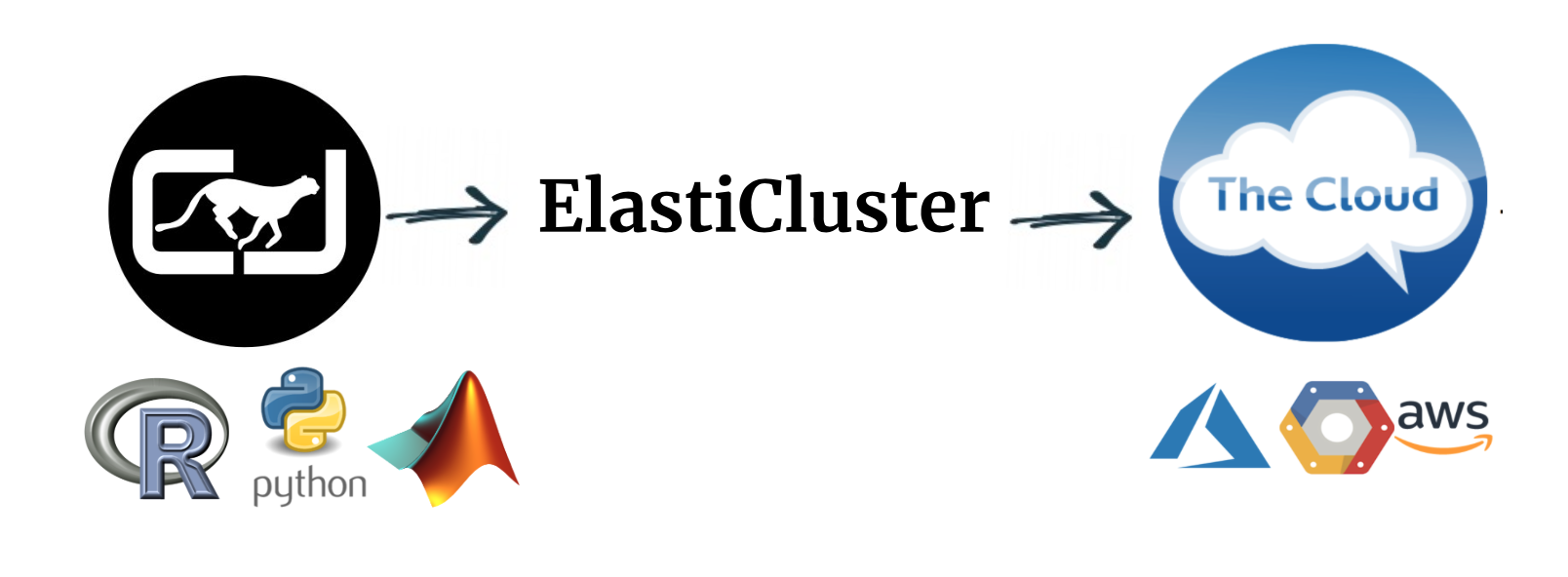}
\caption{
Elasticluster-ClusterJob stack first provisions a personal cluster in the cloud using ElastiCluster, and then links ClusterJob to this cluster to run ambitious experiments involving many parallel jobs. Experiments are documented reproducibly, and can be retrieved at a later time via ClusterJob interface. This stack is agnostic to the choice of cloud provider and programming language.}
\end{figure}

This stack leverages two different components to conduct massive experiments in the cloud: It first provisions services at the \emph{execution} layer
and then exploits services at the \emph{interface} layer to run the actual compute payload.  We are focusing in particular on \emph{ElastiCluster} \cite{elasticluster} to build the virtualized batch-queuing clusters, and \emph{ClusterJob} \cite{clusterjob,MMCEP17} to drive the experiments (see Figure \ref{fig:ec-cj}); We must however emphasize the generality of this model in the sense that the users can use other software systems that offer similar functionalities.

On a more abstract level, this stack includes \emph{building ephemeral clusters} as an additional component of a computational experiment through adopting a Infrastructure as Code (IaC) approach to cloud resource management.
{Currently, the user is responsible to make a call to ElastiCluster to build a cluster, but in the future we expect this step to be handled automatically by ClusterJob.}

This stack was first proposed and implemented by Hatef Monajemi and Riccardo Murri during Stats285 course at Stanford in the Fall of 2017.
Below, we will introduce \emph{ElastiCluster} and \emph{ClusterJob} in more detail.
\begin{itemize}

\item{\bf ElastiCluster.}
ElastiCluster is an open-source software that provides a command line tool and a Python API to create, set up and resize compute clusters hosted on IaaS cloud computing platforms. It uses \emph{Ansible} \cite{Ansible} as an IaC tool to get a compute cluster up and running in a push-button way on multiple cloud platforms, such as AWS, Google Cloud, Microsoft Azure and OpenStack. It offers computational clusters with various base operating systems (e.g., Debian, Ubuntu, CentOS) and job schedulers (e.g., SLURM, SGE, Torque, HTCondor and Mesos). It also supports Spark/Hadoop clusters and several distributed file systems such as CephFS, GlusterFS, HDFS and OrangeFS/PVFS.

\item{\bf ClusterJob (CJ).} ClusterJob is an open-source EMS that makes doing massive reproducible computational experiments on remote compute clusters a push-button affair.
CJ is written in Perl and currently supports batch submission of Python and Matlab jobs to compute clusters via SLURM and SGE batch-queuing systems. For embarrassingly parallel tasks, CJ offers automatic parallelization of scripts that are written serially. In addition, CJ automates reproducibility by generating and saving random seeds for each experiment, list of dependencies and extra code to ensure the results can be reproduced at a later time. Given a main script and its dependencies, CJ produces a reproducible computational package with distinct \emph{P}ackage \emph{ID}entifier (PID) (a SHA-1 code), automatically sets up the execution environment and submits the jobs to a remote cluster. Having the PID, one can track the progress of the runs, harvest the data, and get other information about the experiments at any time using various commands provided by CJ's command line interface.
\end{itemize}

Using this stack and the cloud computing credits that were provided to the students of Stats285 through Google Cloud Education, students were able to setup their own personal GPU clusters in the cloud and collectively train nearly $2000$ deep nets with various architectures and datasets in one calendar day to replicate an important and well-cited article \cite{zhang16} and discover new phenomena in Deep Learning \footnote{The results and discoveries made in Stats285 collaborative study on Deep Learning are expected to be compiled into a peer-review article.}. This model has also been used extensively during 2018 Stats285 Data Science Hackathon\footnote{See course website \url{http://stats285.github.io} } to attack challenging problems in political science, medical imaging and natural language processing. For our own research, each of our members regularly use this model of computing. Our experience shows that it takes roughly 15-18 minutes to setup a CPU cluster with less than 10 computational nodes\footnote{ElastiCluster currently sets up nodes in batches of 10 at a time. So, for a cluster with $N$ node the setup time is roughly $ (1 + \lfloor(N-1)/10\rfloor) \times T$ where $T$ is the setup time for one batch ($T \approx 20$ min).} and 20-23 minutes if GPU accelerators are attached to the nodes (Extra $5$ min is due to time it takes to install CUDA)\footnote{The time it takes to setup a cluster in the cloud can vary slightly due to various factors such as the proximity and network traffic of the cloud provider's data center, responsiveness of the cloud provider API (e.g.
starting a VM on Azure is much more complex than on Google), the boot process of the chosen operating system (e.g., Debian is faster to boot than Ubuntu), the number and speed of CPUs on the local machine, etc.}.


The exact details of this stack are explained thoroughly in the GitHub companion page of this article \cite{companion-page}. The reader is encouraged to setup a personal cluster following the guide therein. We will briefly explain the general idea here.

An individual can spin up a personal HPC cluster (say \verb+gce+) by providing a simple configuration file to ElastiCluster and typing the following command in a terminal:\\

\begin{lstlisting}[frame=single,backgroundcolor = \color{lightgray},basicstyle={\small\ttfamily}, morekeywords={cj, elasticluster}]
$ elasticluster start gce
\end{lstlisting}
here \verb+elasticluster+ is simply a $0$-install bash script that is provided to the user. This script uses a dockerized version of ElastiCluster to execute your command; it pulls ElastiCluster's docker image from DockerHub, and then runs elasticluster in a docker container. If Docker is not installed on your machine, the script asks for your permission to automatically install it.
ElastiCluster's 0-install script thus brings additional convenience to the user by eliminating the need for the installation of ElastiCluster's API and various dependencies.

Once \verb+gce+ cluster is setup, you can run your experiments on it using CJ. All that is needed from your cluster to link it to CJ is the IP address of the frontend (master) node, which can be obtained via the following command:
\begin{lstlisting}[frame=single,backgroundcolor = \color{lightgray},basicstyle={\small\ttfamily}, morekeywords={cj, elasticluster}]
$ elasticluster list-nodes gce
\end{lstlisting}

To use CJ, one has to install it on a local machine. CJ is written entirely in Perl and features a very straight-forward installation guide that is provided in the companion page of this article \cite{companion-page}. Once CJ is available on your machine, you can configure your cluster via either of the following commands:
\begin{lstlisting}[frame=single,backgroundcolor = \color{lightgray},basicstyle={\small\ttfamily}, morekeywords={cj, elasticluster}]
$ cj config gce --update
$ cj config-update gce
\end{lstlisting}

This command prompts the user to setup the new \verb+gce+ cluster by providing the IP address and other optional configuration options such as the desired runtime libraries\footnote{CJ uses conda package manager \url{https://conda.io/docs/} to automatically setup software environment according to libraries determined in {\tt ssh\_config} file.}. The information provided by the user will be saved in CJ's configuration file \verb+~/CJinstall/ssh_config+. For clusters that already exist in this configuration file, a user can update only the corresponding IP address to avoid altering an earlier specification of optional parameters each time a new machine is created.

\begin{lstlisting}[frame=single,backgroundcolor = \color{lightgray},basicstyle={\small\ttfamily}, morekeywords={cj, elasticluster}]
$ cj config-update gce host=35.185.238.124
\end{lstlisting}

After this step, running MCEs on \verb+gce+ is a push-button affair.
As a simple example, consider a Deep Learning experiment (written in PyTorch or Tensorflow) that involves training 50 networks for a grid of $10$ architectures and $5$ datasets. The experimenter first implements a main Python script \verb+DLexperiment.py+ that loops over all 50 $(architecture,dataset)$ combinations and executes a certain task for each. She then includes all additional dependencies including datasets in a directory \-- say \verb+bin/+\footnote{It is also possible to direct CJ to use datasets already on a cluster, hence not moving data from local machine to remote cluster if data will be used for more than one experiment.}. The following CJ command then automatically parallelize \verb+for+ loops inside the main Python script, creates 50 different separate jobs for all the combinations and reproducibly runs them on \verb+gce+ while assigning 1 GPU to each job.

\begin{lstlisting}[frame=single,backgroundcolor = \color{lightgray},basicstyle={\small\ttfamily}, morekeywords={cj}]
$ cj parrun DLexperiment.py gce -dep bin/
 -alloc '--gres:gpu:1' -m 'reminder message'
\end{lstlisting}
Once the computations associated with certain \verb+<PID>+ are finished, the experimenter can harvest the results of all jobs and transfer them to a local machine through various available harvesting commands. As an example, below we reduce all the \verb+results.txt+ files of all jobs into one file and transfer the package to the local machine:
\begin{lstlisting}[frame=single,backgroundcolor = \color{lightgray},basicstyle={\small\ttfamily}, morekeywords={cj}]
$ cj reduce results.txt <PID>
$ cj get <PID>
\end{lstlisting}

The results obtained may then suggest designing and running new experiments, which can be easily handled through CJ. Once all necessary data are collected and the experimenter is satisfied with the current round of experiments, the personal cluster is no longer needed and so it can be destroyed:

\begin{lstlisting}[frame=single,backgroundcolor = \color{lightgray},basicstyle={\small\ttfamily}, morekeywords={cj,elasticluster}]
$ elasticluster stop gce
\end{lstlisting}

The information about the computations conducted through CJ is logged and can be retrieved at a later time. CJ provides a very simple command-line interface (CLI) with many features for managing data science experiments. The reader is referred to CJ's documentation available on \url{www.clusterjob.org} for a comprehensive list of features. It should be emphasized that both ElastiCluster and ClusterJob are open-source software under active development and the reader is encouraged to follow their future enhancements on GitHub.


\subsection{CodaLab Worksheets}
\label{sec:codalab}

CodaLab Worksheets \cite{CodaLab} offer an EMS developed by a team at Stanford University led by Percy Liang and supported by Microsoft.  CodaLab's premise is that in order to accelerate computational research, we need to make it more \emph{reproducible}.  Just as version control systems like Git have enabled developers to scale up software engineering, CodaLab hopes to do the same for computational experiments.  CodaLab allows users to upload code, data, and run cloud experiments.  CodaLab automatically keeps track of the full provenance of computation, so that it is easy to introspect, reproduce, and modify existing experiments.

\begin{figure}
\includegraphics[scale=0.2]{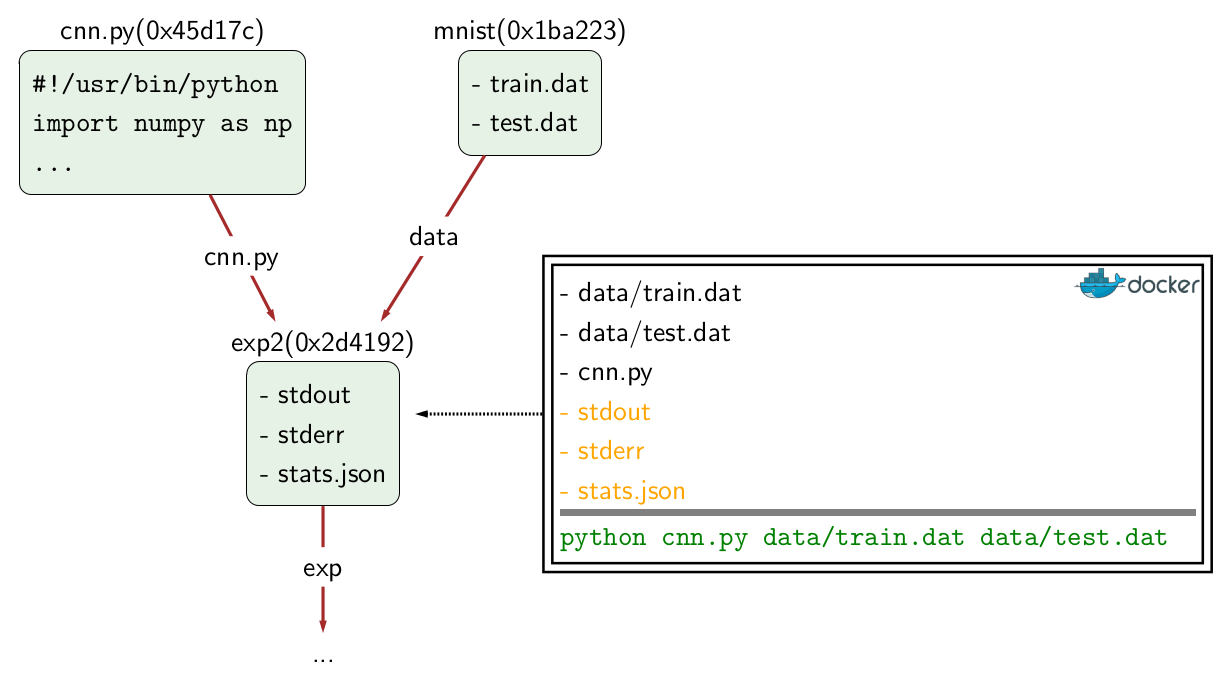}
\caption{\label{fig:codalabExecution}
Execution in CodaLab Worksheets
proceeds by taking a set of input bundles (immutable files/directories representing code or data), running arbitrary code in a docker container,
and producing an output bundle, which can be used further downstream.
}
\end{figure}

CodaLab is built around two concepts\footnote{For more information, visit \url{https://github.com/codalab/codalab-worksheets/wiki}.}: \emph{bundles} and \emph{worksheets}.
Bundles are immutable files/directories that represent the code, data, and results of an experimental pipeline. There are two ways to create bundles. First, users can \emph{upload bundles}, which are datasets in any format or programs in any programming language. Second, users can create \emph{run bundles} by executing shell commands that depend on the contents of previous bundles. A run bundle is specified by a set of bundle dependencies and an arbitrary shell command. This shell command is executed in a docker container in a directory with the dependencies. The contents of the run bundle are the files/directories which are written to the current directory by the shell command (Figure~\ref{fig:codalabExecution}).  In the end, the global dependency graph over bundles precisely captures the research process of the entire community in an immutable way.

Worksheets organize and present an experimental pipeline in a comprehensible way, and can be used as a lab notebook, a tutorial, or an executable paper. Worksheets contain references to a subset of bundles and can be thought of as a view on the bundle graph.  Worksheets are written in a custom markdown language, and in the spirit of literate programming,
allow one to interleave textual descriptions, images, and bundles,
which can be rendered as tables with various statistics.

The CodaLab server takes execution requests and assigns jobs to workers.
The user can view the results (stdout and any files) in real-time and
also communicate with the running process via ports.  One unique property
about CodaLab is that users can also connect their own computing resources
(from one's laptop to one's local compute cluster to AWS Batch) to the CodaLab server, allowing for more decentralization and larger potential
for scaling up organically.

A CodaLab user can either use the public instance (\url{worksheets.codalab.org}) or setup a custom instance (e.g., for a research lab).
CodaLab can be used either from a web interface or from a command-line interface (CLI), which provides experts with more programmatic control.
The CLI is easily installed from PyPI:
\begin{lstlisting}[frame=single,backgroundcolor = \color{lightgray},basicstyle={\small\ttfamily}, morekeywords={pip}]
$ pip install codalab
\end{lstlisting}
This provides the command {\tt cl}, which is the main entry point to CodaLab functionality.

To upload a bundle (either source code or data):
\begin{lstlisting}[frame=single,backgroundcolor = \color{lightgray},basicstyle={\small\ttfamily}, morekeywords={cl}]
$ cl upload cnn.py
$ cl upload mnist
\end{lstlisting}
Recall that bundles can either be files or directories.
To execute an experiment, one must specify the input bundles (in this case, two of them) and a command to be run, producing a run bundle:
\begin{lstlisting}[frame=single,backgroundcolor = \color{lightgray},basicstyle={\small\ttfamily}, morekeywords={cl}]
$ cl run :cnn.py data:mnist \
  'python cnn.py data/train.dat data/test.dat'
\end{lstlisting}
For each input bundle, one specifies a name (e.g., {\tt data}).  The execution of the command
takes place in a Docker image where the input bundles are presented as files/directories
with the given names in a temporary directory.  The command outputs additional files in the
current directory, which are saved as the contents of the run bundle once the bundle finishes
executing.

The CodaLab execution model is based on dataflow, in which bundles represent information processed in a pipeline.  In particular, bundles are immutable, so each command produces a new run bundle
rather than modifying an existing bundle.  Note that the Docker environment is only used
temporarily to run the command; only the outputs of the run are saved.
This immutability stands in contrast to other execution models where one might have an
entire virtual machine at one's disposal or in the case of Jupyter notebook,
the entire Python kernel.  The dataflow model of CodaLab is important for introspection
and decentralized collaboration: one can see exactly the chain of commands that were run,
and another researcher can build off of an existing result by simply running more commmands
on it without a danger of overriding anything.

One of the most powerful features in CodaLab is called {\tt mimic}, which is enabled by
having the dataflow model of computation.  In brief, {\tt mimic} allows you to rerun
a computation with modifications.  The basic usage is as follows:
\begin{lstlisting}[frame=single,backgroundcolor = \color{lightgray},basicstyle={\small\ttfamily}, morekeywords={cl}]
$ cl mimic A B
\end{lstlisting}
This command examines all the bundles downstream of {\tt A}, and re-executes them all,
but now with {\tt B} instead.
For example, {\tt A} could be the old dataset and {\tt B} could be the new dataset, or {\tt A} could be the old algorithm and {\tt B} could be the new algorithm.
In principle, whenever someone creates a new method, she should be able to painlessly
execute it on all existing datasets as long as the new method conforms to a standard interface.

One of the two main uses of CodaLab Worksheets is in running the Stanford Question Answering Dataset (SQuAD) competition (\url{stanford-qa.com}).  In this competition, researchers have to develop a system that can
answer factual questions on Wikipedia articles.  Over the last two years, over 70 teams
have submitted solutions to the highly competitive leaderboard.  Each team runs their models
on the development set, which is public; this allows teams to independently configure their
own environment and manage their own dependencies.  Once the team is ready, they submit their
system, which is actually the bundle corresponding to the result on the development set.
The SQuAD organizers use {\tt cl mimic} to re-run the experiment on the hidden test set instead.
Here, we see that CodaLab provides both flexibility and standardization.
As a case study, in the past two years, one of the homeworks in the Stanford Natural Language Processing class has been to develop a model for SQuAD.  In 2017, 162 teams from the class participated using the public instance of CodaLab, which was able to scale up and handle
the load.

The other main use case of CodaLab is to help people run and manage many experiments at once.
This happens at several levels: First, CodaLab is backed by a cluster, so that the user needs
to only focus on what experiments to run rather than where to run them (although the user
can specify resource requirements).  Second, the dataflow model means that for every experiment,
which version of the code and data used to produce that experiment is fully documented;
as a result, one will never find oneself in a situation with a positive result that is not reproducible anymore.  Third, the worksheets in CodaLab offer a flexible
way of monitoring and visualizing runs.  One can define a table schema, which specifies the custom fields to display (e.g., accuracy metrics, resource utilization, dataset size); a run is
a row in this table.  This allows one to easily compare the metrics on many variants
of the same algorithm, leading to faster prototyping.

In summary, CodaLab provides a collaborative platform that allows researchers to contribute
to one global ecosystem by uploading code, data, and other assets, running experiments
to generate other assets (bundles), etc.  CodaLab keeps the full provenance and provides full transparency (though one can opt to keep some bundles private if necessary).
CodaLab starts as a mechanism for enabling researchers to be more efficient at running experiments, and also serves as a publishing platform for published research or competitions.
Having a common substrate that supports these use cases opens up the opportunity to bring
development and publication closer together.

\subsection{PyWren's serverless execution model}
\label{sec:transparent-invocation-model}

\begin{table*}
\caption{Serverless computing resource limits per invocation}
\centering

\begin{threeparttable}
\begin{tabular}{l|ccc}
\hline
 & AWS Lambda\tnote{1} & Google Cloud Functions & Azure Functions \\
\hline
Deployment (MB) & $50$ & $100$ & N/A  \\
Memory (GB) & $3.0$ & $2.0$ &  $1.5$ \\
Ephemeral Disk (GB) & $0.5$ & $2.0-Mem$\tnote{2} & 5000\\
Max. run time (sec) & $300$ &  $540$& $600$
\end{tabular}
\begin{tablenotes}
\item[1]{\scriptsize \url{https://docs.aws.amazon.com/lambda/latest/dg/limits.html}}
\item[2]{\scriptsize Disk space consumes from the memory limit.}
\end{tablenotes}
\end{threeparttable}
\label{tbl_serverless_limits}
\end{table*}

Many scientific computing tasks exhibit a significant degree of innate parallelism, which if properly exploited can dramatically accelerate computational science. These range from classic Monte Carlo methods, 
to the optimization of hyperparameters, 
to featurization and preprocessing of large input volumes of data. In many of these cases, large amounts of code are written a priori for one \emph{instance} of such tasks without regard to potential parallel or distributed execution. Indeed, it is only at the end (i.e., the outer per-instance processing loop) that parallelism is even apparent.
The code written in this way is called \textit{embarrassingly parallel} because it is trivial to execute in parallel; each operation does not depend on the result of any other. If the computing resources available were truly infinite, the total runtime for these operations would be bounded by the duration of the slowest single scalar piece. Running a single task and running 10,000 tasks would take the same amount of time.

PyWren \cite{Jonas17} is a system developed to enable this kind of massively-parallel, transparent execution.  PyWren is built in the Python programming language, and exploits the language's inherent dynamism to transparently identify dependencies and related libraries and marshal them to remote servers for execution. It uses recent \emph{serverless} platforms offered by cloud providers to quickly command controls of tens of thousands of CPU cores, run the resulting parallel task transparently, and then shut down those machines.

Serverless computing (a.k.a. Function as a Service (FaaS)) is a fairly new cloud execution model in which the cloud provider removes much of the complexity of the cloud usage by abstracting away server provisioning \cite{miller15}. In this model, a function and its dependencies are sent to a remote server that is managed by the cloud provider and then executed. AWS Lambda, Google Cloud Functions and Azure Functions are amongst popular serverless compute offerings.

Current serverless computing services are suitable for short-lived jobs with small storage and memory requirements because of the limits set by the cloud providers (See Table \ref{tbl_serverless_limits}). This is because serverless computing is originally designed to execute event-driven, stateless functions (code) in response to triggers such as actions by users, or changes in data or system state. Nevertheless, serverless computing provides an efficient model for applications such as processing and transforming large amount of data, encoding videos, and applications such as simulations and Monte Carlo method with large innate amounts of parallelism \cite{Jonas17,Ishakian18}.

PyWren can easily be installed via PyPI and following a number of setup prompts which involve providing credentials for authentication to the underlying cloud computing provider:

\begin{lstlisting}[frame=single,backgroundcolor = \color{lightgray},basicstyle={\small\ttfamily}, morekeywords={cl}]
$ pip install pywren
$ pywren-setup
\end{lstlisting}

As an example, consider the following \texttt{MatVec} function that performs the relatively trivial task of generating a random matrix and vector from a $\mathcal{N}(0, b)$ distribution and computing their matrix-vector product, and returning the result.
\begin{lstlisting}[frame=single,backgroundcolor = \color{lightgray},basicstyle={\small\ttfamily}, morekeywords={cl}]

 def MatVec(b):
    x = np.random.normal(0, b, 1024)
    A = np.random.normal(0, b, (1024, 1024))
    return np.dot(A, x)

\end{lstlisting}

Using PyWren's \texttt{map} command, one can painlessly invoke 1,000 distinct instances of this function to be executed transparently in the cloud:

\begin{lstlisting}[frame=single,backgroundcolor = \color{lightgray},basicstyle={\small\ttfamily}, morekeywords={cl}]

  pwex = pywren.default_executor()
  res  = pwex.map(MatVec,
             np.linspace(0.1, 100, 1000))

\end{lstlisting}

Behind the scenes, PyWren exploits Python's dynamic nature to inspect all dependencies required by the function, and marshals as many of those as possible over to the remote executor. The resulting function is run on the remote machine, and the return value is serialized and delivered to the client.

The dynamic, language-embedded nature of PyWren makes it ideal for exploratory data analysis from within a Jupyter notebook or similar interactive environment. 
PyWren is currently limited to exploiting \texttt{map}-style parallelism, although active research is underway to broaden the capabilities of the serverless execution model. The function serialization technology is not perfect -- currently it struggles with Python modules which have embedded C code, requiring them to be packaged independently as part of a runtime. This too is an active area of research. Finally, the limitations provided by the cloud providers' serverless execution environments (including runtime and memory) constrain the exact functions that can be run, although we anticipate these constraints lessening with time.

\subsection{Third-Party Unified Analytics Interfaces}
Several companies provide paid services for painless computing in some third-party cloud; researchers may choose to use their services for conducting their ambitious experiments. A few examples of such companies are Databricks \cite{databricks}, Domino Data Lab \cite{domino}, FloydHub \cite{floydhub} and Civis Analytics \cite{civis}. Each of these companies have a slightly different focus (e.g, Databricks focuses on Spark applications whereas FloydHub focuses on Deep Learning) and may use a different computing model for managing computations in the cloud. Nevertheless, all of them build wrappers around the cloud so that individual users can conduct their MCEs without having to directly interact with the cloud. They provide graphical user interfaces through which users can setup their desired computational environment, upload their data and codes, run their experiments and track their progress. They also offer a community edition of their services that can be used for initial testing before buying their computing and storage services.


\section{Concluding Remarks}
We have presented several computing stacks that can be used to dramatically scale up computational experiments, painlessly. Such stacks constitute what we have called experiment management systems, a fundamental concept in modern data science research. They offer efficiency and clarity of mind to researchers, by organizing the specification and execution of large collections of experiments, removing the apparent barriers to using the cloud. In addition they painlessly capture and document the numerous iterative attempts that get tried in typical ambitious research. 

We look forward to a future where \emph{every} researcher can dream up ambitious computational experiments, open up his/her laptop, and command a computational agent to fire up millions of jobs to study a certain problem of interest.
A future where instead of manual human intervention, computational agents seamlessly run jobs in the cloud, manage their progress, harvest the results of the experiments, run specified analyses on those results and package them in a unified format that is transparent, reproducible and easily sharable. Such automation of research activities will, we believe, empower data scientists to deliver many more breakthroughs and will accelerate scientific progress.


\section{List of Abbreviations}
\noindent 
{\bf AWS} Amazon Web Services\\
{\bf CJ} ClusterJob\\
{\bf EMS} Experiment Management System\\
{\bf GCP} Google Cloud Platform\\
{\bf HPC} High Performance Computing\\
{\bf IaaS} Infrastructure as a Service\\
{\bf IaC} Infrastructure as Code\\
{\bf MCE} Massive Computational Experiment\\
{\bf NMT} Neural Machine Translation\\
{\bf PaaS} Platform as a Service\\
{\bf PCS} Painless Computing Stack\\
{\bf SaaS} Software as a Service\\
{\bf SLURM} Simple Linux Utility for Resource Management\\

\section{Acknowledgements}
This work was supported by National Science Foundation grants DMS-0906812 (American Reinvestment and Recovery Act),
DMS-1418362 (`Big-Data' Asymptotics: Theory and
Large-Scale Experiments) and DMS-1407813 (Estimation and Testing in Low Rank Multivariate Models).
We thank Google Cloud Education for providing Stanford's Stats285 class with cloud computing credits.
We would like to thank Eun Seo Jo for helpful discussions.
\bibliographystyle{apacite}
\bibliography{ref}
\end{document}